# Simulation-based Modelling of Growth and Pollination of Greenhouse Strawberry

Zhihao Cao, Hongchun Qu*

*College of Information Science and Engineering, Zaozhuang University, Zaozhuang 277160, China*

## Abstract

The cultivated strawberry *Fragaria × ananassa Duch.* is widely planted in greenhouses in China. Its production heavily depends on pollination services. Compared with artificial pollination, bee pollination can significantly improve fruit quality and save considerable labor requirement. Multiple factors such as bee foraging behavior, planting pattern and the spatial complexity of the greenhouse environment interacting over time and space are major obstacles to understanding of bee pollination dynamics. We propose a spatially-explicit agent-based simulation model which allows users to explore how various factors including bee foraging behavior and strawberry phenology conditions as well as the greenhouse environment influence pollination efficiency and fruit quality. Simulation experiments allowed us to compare pollination efficiencies in different conditions. Especially, the cause of bee pollination advantage, optimal bee density and bee hive location were discussed based on sensitivity analysis. In addition, simulation results provide some insights for strawberry planting in a greenhouse. The firmly validated open-source model is a useful tool for hypothesis testing and theory development for strawberry pollination research.

## 1. Introduction

Strawberry is any of the various, low-growing perennial plants of the genus Fragaria in the rose family. The most common strawberries grown commercially are cultivars of the garden strawberry *Fragaria × ananassa Duch.,* which has been the most widely distributed fruit crop in the world. It is grown in every country with a temperate or subtropical climate and even in many tropical countries in highland areas, where the climate is mild. Strawberry fruits are highly prized for their universal appeal to the human senses of sight, smell, and taste [9].

Strawberries can be pollinated by self, wind, and bees. Honey bees (*Apis mellifera ligustica*) are recognized as the main pollinator of the strawberry crop [16]. If there are no honey bees, the combined action of gravity and wind assures the basis of the pollination because the stamens can scatter pollen onto many of the pistils as they crack open [13]. However, these flowers may not be completely self-fertilizing in this way, which may result in a low fruit setting rate and malformed fruits. In practical strawberry planting, bees are important for pollination. Pollination by bees not only increases crop yield, but also improves aspects of fruit quality, including nutritional value and shelf life [34]. Recent research suggested that when strawberry plants in greenhouses were isolated from honeybees, the fruit set was 50–59% lower compared to the case where bees were present, which can achieve on average 80% of fruit set when bees were present [16,40,41].

Understanding the bee pollination process is rewarding for improving strawberry yield and quality [31,41,48]. However, many factors such as bee foraging behavior, planting pattern and the spatial complexity of greenhouse environment interacting over time and space are major obstacles to understanding of bee pollination dynamics. For example, bees are attracted by the appearance of plant inflorescence during foraging, so the foraging patterns might differ in various landscapes [45]. This suggests that planting pattern in a greenhouse could affect bee foraging distance and direction and consequently result in the dynamics of pollination efficiency. This interaction in space is difficult to understand quantitively without the help of simulation models, because bee behavior and plant floral exhibit obvious spatial heterogeneities and vary across individuals [46]. Previous strawberry models focused on strawberry growth processes [22] including modelling the water-balance for irrigation [49], statistical yield forecasting via weather conditions [50], and prediction of phenological stages based on regressions [51]. These models do not include pollination processes. Modelling the spatial and stochastic process, spatial-explicit individual based models are much more intuitive and powerful than conventional ordinary differential equations.

Hence, we modelled the strawberry plants and honey bee foraging behavior for the first time and use quantitative methods to analyze simulation results of fruit production. Modelling strawberry pollination allows us to decipher hidden relationships between variant organisms to better understand the pollination service in greenhouse strawberries. This model was used to study the bee pollination process in a strawberry greenhouse. Specifically, we studied the influence of bee density on strawberry quality, the cause of advantage for bee pollination over artificial pollination, and the influence of hive location on strawberry quality. We proposed some planting suggestions for strawberry growers based on the experiment results. The main contributions of this paper are as follows:

(1) We propose an open-source model that provides multi-scale simulation from strawberry pollen to plant, including simulation of strawberry growth in a greenhouse and interaction with bees. This model incorporates much of what is known and hypothesized about the pollination ecology of strawberry agroecosystem. The model codes include many adjustable parameters that allow users to adapt to different cultivars and planting environments.

(2) We analyzed the optimal bee density based on the proposed model. The results suggested that when the bee density was higher than 1.00 bee/clone, the fruit quality was not significantly improved due to a saturation effect. In practical greenhouse planting, strawberry pollination may be affected by mechanical actions, diseases and other factors, with a result that it is necessary for growers to ensure the bees outnumber the strawberry clones in a greenhouse.

(3) The effect of hive location and bed spacing in a greenhouse on pollination efficiency was analyzed based on the model. We suggested that strawberry growers can place bees in multiple hives not only one hive and then place these hives in different locations in a greenhouse to diminish the influence of bee foraging distance constraints.

(4) Practical planting experience suggests that bee pollination is better than artificial pollination in many respects. For the first time, the quantitative effects of viability of pollen and stigma, self-compatibility on strawberry fruit quality were analyzed through simulation experiments. The results revealed that the even distribution of pollen in pistil during bee pollination is the primary cause and stigma receptivity is a secondary cause of advantage for bee pollination over artificial pollination. Not only the pollen transport but the even pollen distribution caused by bee behavior are significantly important.

## 2. Materials and methods

### 2.1 Strawberry

The cultivated strawberry *Fragaria × ananassa Duch.* is widely planted in greenhouses in China [41]. Therefore, we selected this cultivar because of its extensive history of study [4,8,11,13,21,34] and economic importance. Specifically, its bee pollination process was studied through simulation-based modelling. For other cultivars, the simulation software codes include a variety of adjustable parameters on strawberry growth that users can modify to adapt to different cultivars of strawberry.

### 2.1.1 Greenhouse

The most popular strawberry cultivation system in China is solar-powered plastic greenhouse and this type of cultivation system uses only solar energy for crop production, as shown in Figure S1. The greenhouse not only prevents the damage caused by adverse climatic conditions but also provides a suitable environment for strawberry cultivation and protects the crop from insects and pets [32]. The microclimate in greenhouses is significantly more suitable than that in open environment.

Strawberry fruit production in China generally takes place from November to April, starting with the planting of bareroot transplants in November. In this simulation, it is assumed that the cultivation site is in China and the simulation starts on January 1st and lasts until April, lasting around 120 days. There is a close relationship between solar light and inside temperature in a greenhouse. All the greenhouses were managed by experienced fruit growers and were controlled at a fixed temperature and relative humidity (RH) condition [1]. Low temperatures will increase the possibility of damaged fruits as well as changes in fruit size and high temperatures reduce the plant's photosynthetic rate [33]. Among the environmental conditions, temperature is one of the most important factors [44] that affect fruit and seed set at different stages of the reproductive developing process. The general temperature ranges [15,32] in two weather conditions and months are shown in Table 1.

Table 1 The setting of general temperature range of the greenhouse in simulation

| Month | Temperature range (sunny) | Temperature range (cloudy) | Day of simulation |
|---|---|---|---|
| January | 6-22°C | 6-12°C | 1-30 |
| February | 10-26°C | 12-14°C | 31-60 |
| March | 14-28°C | 13-16°C | 61-90 |
| April | 16-30°C | 14-18°C | 91-120 |

Standard greenhouses in China are usually 80 m long and 8 m wide. Strawberry plants were grown in raised beds with a plastic film cover on the soil (plastic mulching). Distance between two strawberry beds is usually about 0.4 m, and the single bed width is about 0.6 m. There are two rows of strawberry clones per bed, about 0.24 m apart. The distance between two strawberry clones is about 0.2 m. As a result, there are 12 rows in the greenhouse and 390 strawberry clones in each row. Totally, this standard greenhouse includes 390*12*2=9,360 strawberry clones. The bee hive is generally arranged on the east side of the greenhouse where the temperature is high according to practical planting experience. The simplified greenhouse diagram is shown in Figure 1.

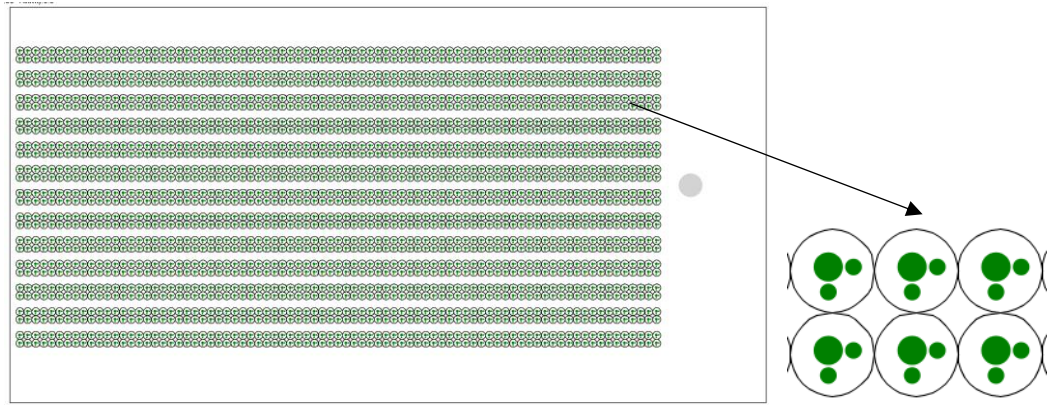

Fig.1 A simulated strawberry greenhouse. The out rectangle represents the greenhouse boundary. Inside the boundary, black open circles represent strawberry clones, green dots represent different inflorescences. Green dots of different sizes represent different inflorescence rank. In the east open site of the greenhouse, the grey dot represents bee hive.

### 2.1.2 Inflorescence and flower

Strawberry flower clusters occur on a series of double branches with a flower in the fork of each branch [45]. Generally, strawberry plants have one primary flower, two secondary flowers and up to four tertiary flowers per inflorescence. The existence of competition within inflorescences was demonstrated by obtaining heavier secondary berries after removing the primary [11]. Therefore, only primary and secondary fruits were used to estimate yield, as they are the only fruits usually considered marketable [11]. In order to improve yield and quality of strawberry fruits, it is usually necessary to cut off some low-level inflorescences in production, i.e., flower thinning. In our simulation, each strawberry clone retains one primary inflorescence and two secondary inflorescences. There are of 6 flowers in each primary inflorescence and 3 in each secondary inflorescence in simulation, as shown in Table 2. Therefore, a strawberry clone can bear up to 12 fruits.

Table 2. Flower information of strawberry in simulation

| Inflorescence | Number per clone | Number of flowers per inflorescence | Number of pistils(ovules) per flower | GDD* |
|---|---|---|---|---|
| Primary | 1 | 6 | 350 | N(586, 70) |
| Secondary | 2 | 3 | 260 | N(946, 70) |

*: The Growth Degree Day for strawberry plants is assumed to follow normal distribution.

Strawberry fruit weight and achene number showed a consistent decrease from primary flowers to the secondary ones. The primary flowers bear about 350 stigmata (ovules), the secondary ones about 260, and the tertiary ones only about 180 [4]. The pistils are arranged in a regular spiral on the stem end of the receptacle. The pistil base commonly called the seed, contains one ovary. One ovary contains one ovule, and each achene can be considered as a single seed that is actually an individual fruit, called an achene [13].

The strawberry flowers bloom in sequential order starting with primary flowers, followed by secondary flowers and finally up to tertiary flowers [35]. The primary flowers bloom earlier and secondary flowers bloom later with a time gap of around 20 days [8,21,34] with a result that there are usually two harvest peaks in continuous harvests. The average weight of berries produced decreased at each harvest, which is a common experience for strawberry growers [27]. A flower contains around 25 stamens (anthers) and a large number of strawberry pollen grains. Generally, each anther contains about 18,000 normal pollen grains and a flower contains 1800*25=45,000 pollen grains.

### 2.1.3 Strawberry fruit weight

These achenes, resulting from fertilized ovules, are large and surrounded by well-developed fleshy tissue, whereas the achenes resulting from unfertilized ovules are less voluminous and closer together. The weight of a berry is proportional to the number of fertilized ovules (achenes), the number of stigmata per flower determines its potential weight [2]. Primary berry is the largest and that the secondary, tertiary, and quaternary are gradually smaller. Secondary berry is about 80 percent of the primary [13]. Therefore, pollination of all pistils of the flower are necessary for maximum berry size.

Assuming the fruit is evenly developed and there is a fitted linear equation between fruit weight and total achene number [11,35] which is derived from the observed data:
$$Y = 0.05 * X + 2.0$$
Where $Y$ is the weight of the strawberry fruit (g) and $X$ is the number of achenes per strawberry fruit [4]. Adjustments have been made to the equation according to the contemporary strawberry empirical data collected in [5], as shown in Figure S5. It should be noticed that the fitted equations for variant strawberry cultivars are not the same.

### 2.1.4 Malformed Fruit

Fruit malformation is a phenomenon commonly observed in commercial strawberry cultivars with a negative impact on the economic benefit of crop production [36]. Fruit shape, when not damaged by diseases, frost or mechanical action [38], depends on the percentage of fertilized ovules [30]. Malformed fruits have significantly fewer achenes than well-formed fruits, and unfertilized achenes lead to malformation. Successful fertilization and adequate numbers of achenes are necessary for a marketable shape and size of strawberry fruits [47]. The environment in a greenhouse is generally suitable for strawberry growth, so only the influence of pollination is considered in the simulation while other environment factors are ignored.

Generally, artificial pollination results in a low fruit-setting rate, high malformed fruit rate, poor fruit shape and consequently low economic benefit in strawberry planting. Bee pollination can not only improve fruit-setting rate, but also improve fruit quality and reduce malformed fruit rate [16]. There is no universal explanation for bee pollination advantage.

An average of 11 honey bee pollinator visits per flower is required to attain normal berry (fertilization rate, 87%) [6]. Achenes containing fertilized ovules can release a hormone that stimulates growth of the receptacle. When an achene does not contain a fertilized seed, it remains small and the receptacle in its area fails to grow [27]. Only fruits with a high percentage (more than 87%) of fertilized achenes will develop normally without major malformations that reduce overall yield and marketability [12]. Fruits are deformed at parts in which achenes are not fertilized. Malformed fruits are common during strawberry production, with an average rate of about 3%-10% [15,16].

### 2.1.5 GDD

Thermal time models are widely used in vegetation phenology, and they are based on the accumulation of growing degree days (GDD) or heat units. The GDD methodology consists of defining a constant base temperature and then calculating the sum of those temperatures that exceed the base temperature $T_{base}$ in the temperature time series.

The parameters for the strawberry development GDD model are based on reference [7, 44]. In this simulation, the daily cumulative GDD formula is simplified as:

$$GDD = (T_{max} + T_{min})/2 - T_{base}$$

Where $T_{max}$ is the daily maximum temperature，and $T_{min}$ is the daily minimum temperature. In this model, it is assumed that the simulation starts from January 1st for simplicity, strawberry clones enter the blooming period in early February, and enter the peak of fruit drop in early March. Based on the reference [7, 44], the GDD requirement follows the Gaussian normal distribution with the mean of 586. Monte Carlo simulation method and empirical data are used to estimate the standard deviation which can be set to about 70.

The blooming time in different inflorescences is also different. Normally the blooming time of primary flowers is earlier than that of secondary flowers with an average interval of about 20-30 days. In this model, it is assumed that the required accumulated GDD distribution for primary flowers follows N(586, 70), and the distribution for secondary flowers follows N(946, 70), as shown in Table 2. Therefore, there are two harvest peaks i.e., first harvest and second harvest due to the blooming time gap [7]. The accumulated GDD requirement for ripeness follows N(284, 30) with $T_{base}$ of 3°C. Ripening of strawberry fruits is characterized by red coloring and softening. As the temperature when secondary flowers bloom is higher than primary flowers, the time for secondary fruit ripeness is shorter. Average time for single fruit development in first harvest is about 30 days, while that in second harvest is about 24 days, as shown in Table 3. Timing schedule for two harvests and two blooming peak periods are shown in Figure 2.

Table 3. Accumulated GDD requirements information

| Harvest | Inflorescence rank | GDD sum | Time for ripeness |
|---|---|---|---|
| First | primary | N(284, 30) | About 30 days |
| Second | secondary | N(284, 30) | About 24 days |

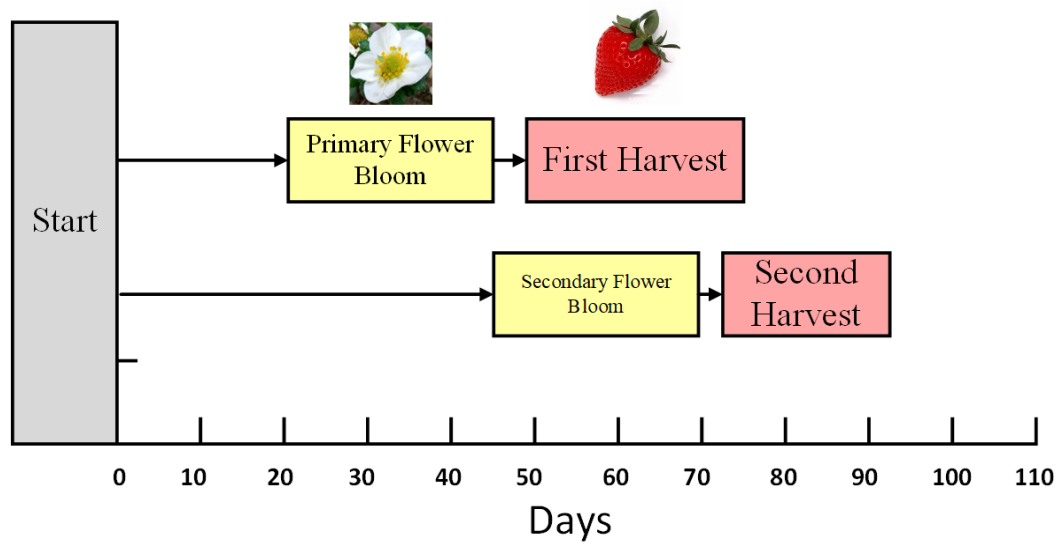

Fig.2 Timing schedule for the two harvests of strawberry simulation

## 2.2 Honey Bee

Most strawberry anthers begin to dehisce after the flowers have bloomed. In open fields, shaking by wind is usually sufficient to trigger this pollen release, promoting self-fertilization for most crops. In the absence of wind, however, as is the case in greenhouses, successful wind pollination of strawberry flowers is difficult [20]. Strawberries in a greenhouse can be pollinated by artificial methods or pollinators. Some research suggests that strawberries with artificial pollination tend to have a low fruit-setting rate and high malformed-fruit rate [16], while bee pollination can significantly improve the fruit-setting rate, fruit yield, fruit quality and saved labor requirements [42].

Currently, 80% of strawberry greenhouses in China use the Italian honeybee (*Apis mellifera ligustica*) which is highly commercialized for pollination [16,17,24]. It is the most common bee species in China. The number of bees is commonly determined by growers based on the greenhouse area and the strawberry planting density. There are about 4000-8000 bees in one colony [31], but the hive-leaving rate (number of bees leaving the nest for foraging/ total bee number in the hive) is low [23]. Hierarchy exists within the bee colony, and the colony is composed of queen bees, drone bees and worker bees. Worker bees can be divided into bees doing internal service and bees doing external service according to the age. Generally, only about 40% of the worker bees actually forage outside in the greenhouse every day [18] and this proportion is influenced by population structure and food factors.

### 2.2.1 Bee Activity

Many studies suggest that weather conditions affect the foraging behavior of honey bees, of which temperature is the most important [1,19]. The suitable temperature range for bees to forage is 15-25 °C, and bees will stop foraging when temperature is below 14 °C or above 30 °C [16]. Affected by temperature and light, bees are basically inactive in cloudy days, and only occasionally active around noon when the temperature is high in a day [14].

Bee foraging activity is set to a period from 8:00 to 17:00 every day [3, 6]. In the morning and evening, bees are basically inactive, while in the hot noon, bees are active with a peak from 11:00 to 14:00 [3]. The activity of bees is mainly affected by the temperature in the greenhouse and can be described by the proportion of bees leaving nest for foraging at this time period [14, 15, 31], as shown in Table 4 and Figure S6. It should be noted that some worker bees may not go foraging due to age. For simplicity, it is assumed that hive releases bees on the hour scale in the simulation.

Table 4. The proportion of bees leaving the nest on different hours

| Time | Ratio (sunny) | Ratio (cloudy) | Time | Ratio (sunny) | Ratio (cloudy) |
|---|---|---|---|---|---|
| 8:00 | 0 | 0 | 9:00 | 0 | 0 |
| 10:00 | 1.13% | 0 | 11:00 | 2.03% | 0 |
| 12:00 | 2.70% | 0.25% | 13:00 | 3.15% | 0.50% |
| 14:00 | 2.60% | 0.25% | 15:00 | 1.20% | 0 |
| 16:00 | 0 | 0 | 17:00 | 0 | 0 |

During foraging, bees are capable of controlling their flight speed [25] in searching flowers to achieve a tradeoff between searching efficiency and accuracy [26]. In the simulation, a bee can complete one activity every minute including searching for nearby flowers, visiting flowers, and returning to the hive [10]. The process is scheduled by phenology and physiological conditions in the greenhouse.

### 2.2.2 Bee-Plant Interaction

Honey bees always landed on the top of flowers and performed a circular movement around the flower, allowing contact with basal pistils during visits. Bees mainly spread pollen grains by body contact. The flower visiting pattern is centralized and adopts the principle of proximity with continuity and repetition in flower visiting. Normally bees can visit several flowers in succession on the same inflorescence. After visiting all flowers in one inflorescence, bees will fly to other inflorescences nearby to continue visiting. Bees move infrequently among inflorescences, with an average distance of 1.1m. According to the references [14, 23], the average number of flowers visited by Italian bees per minute is 2.5-3.8, the average interval between visits is about 10s, and the residence time of a single flower is about 10s.

Flower pollination must be successful after 11 honeybee visits [4], and the number of deposited pollens for a honeybee per visit is about 30 [15], which is consistent with existing literature [6, 10]. The pollen removal ratio is about 20% [15], so the number of pollen grains removed per visit is about 9000. The pollen number in these two activities is significantly different because the pollen collected by bees exists in the form of clusters with strong viscosity, which are formed by a large number of pollen grains bonded by nectar and bee saliva.

It is usually agreed that pollen availability might be a factor in determining which flower to visit. However, as for strawberry flowers, reference [14] suggests Italian bees are not selective in strawberry flowers and the probability of visiting flowers of different ages is basically equal.

## 2.3 Strawberry Pollination

In a strawberry greenhouse, growers can brush the outer edge of the flower with a paintbrush to move pollen into the center of the same flower, which is called artificial pollination. Compared with artificial pollination, bee pollination results in better strawberry quality.

Pollination processes mostly happen during the first 4 days after the flowers bloom [39]. In the model, the blooming period of strawberry is set to 5 days. The suitable temperature for pollen germination is 20-30 °C, which is similar to greenhouse temperature. The proportion of dehisced anthers increased with increasing temperature and peaked at noon, coinciding with the peak pollen-collecting activity of bees [1].

Fertilization rate of strawberry fruits that are marketable must amount to 87% in which the rate is calculated for each fruit as the percentage of achenes with fertilized ovules in relation to the total number of achenes [30]. After the deposited pollens successfully falls on the stigma through honeybee activity, the ovule fertilization process comprises two stages: receiving stage and accepting stage.

### 2.3.1 Receiving stage

At receiving stage, the probability of receiving pollen grains of a stigma positively correlates with pollen viability and stigma receptivity [1]. Viability of the pollen deposited by bees onto stigmas is an important component of individual pollination effectiveness [28]. Stigma receptivity is mainly determined by flower age, and the viability is peak within three days of blooming and sharply declines in 5 days [43], as shown in Figure S17. Value $R$ ranged 0-1 can be used to quantify the stigma viability, as follows:

$$R = e^{-0.01*age^{3.6}} \quad (1)$$

On the other hand，not all of pollen grains carried by bees have high viability. Related research shows considering different periods comprehensively, the average proportion of pollen with high viability is about 50% -70% [38,43], while the mixed saliva of bees when collecting pollen may reduce the viability [28] with a result that proportion of active pollen carried by Italian bees is about 30% [23]. Therefore, the pollen receiving rate in this stage can be described as follows:

$$P_{\text{receive}} = 0.3 * e^{-0.01*age^{3.6}} \quad (2)$$

### 2.3.2 Accepting stage

Each strawberry flower contains both the male stamen and the female pistil. It is generally agreed that strawberry flowers are hermaphroditic and self-compatible to a certain extent, but not complete.

At accepting stage, pollination process is mainly influenced by strawberry acceptance (acceptance rate = mean compatible pollen grains/mean received pollen grains). Hence even if a pollen grain has high viability and the sigma has high receptivity, it may not necessarily lead to pollen tube growth. Noted that the degrees of self-compatibility in different cultivars are not the same [45] and there is no correlation between pollen viability and pollen self-compatibility.

When strawberry growers use self-pollination in practical planting, native pollen grains fall on the pistil of the same flower, which lead to fruits of low quality. However, these flowers pollinated with both foreign and native pollen grains by pollinators can develop large fruits of high quality [37]. These flowers using supplementary pollen from nearby strawberry flowers have better fruit quality compared with those with only native pollen [34]. More than four strawberry cultivars benefited from cross-pollination [44]. Self-incompatibility can prevent fertilization by an individual plant's own pollen, promoting heterozygosity and the long-term adaptability of populations [44]. The interaction of cognate alleles in strawberry prevents successful pollen growth in the self-fertilization. Comparing the number of strawberry bee fruit seeds and fruit weight under different pollination patterns, it is assumed that the pollen self-compatibility probability is about 80% preliminarily [5,37,44] and sensitivity analysis will be conducted on it in following sections to verify its influence.

## 2.4 Modelling

### 2.4.1 Modelling logic

We modelled strawberry pollination by identifying the significant entities and interaction processes of entities, and then scheduling these agents based on empirical data. The simulation model consists of two types of significant entities: (1) regular entities, e.g., greenhouse, strawberry clones, inflorescence and flowers and honey bees in the real-world production system; and (2) virtual entities, e.g., the environment, system scheduling, weather that provide a spatial and temporal reference for the interactions among regular entities [10].

The entities are modelled as agents interacting with each other in a virtual greenhouse, as shown in Figure 3. The greenhouse is composed of 9,360 strawberry clones. One primary and two secondary inflorescences with flowers are distributed within one clone. A fixed number of flowers is reserved by flower thinning according to the inflorescence rank. Bees are spatially initialized in hive site that is arranged on the east side of the greenhouse. Bee foraging is modelled as searching and flower visiting from one flower to another. The pollination interacting process between flowers and bees is defined as pollen exchange with receiving stage and accepting stage.

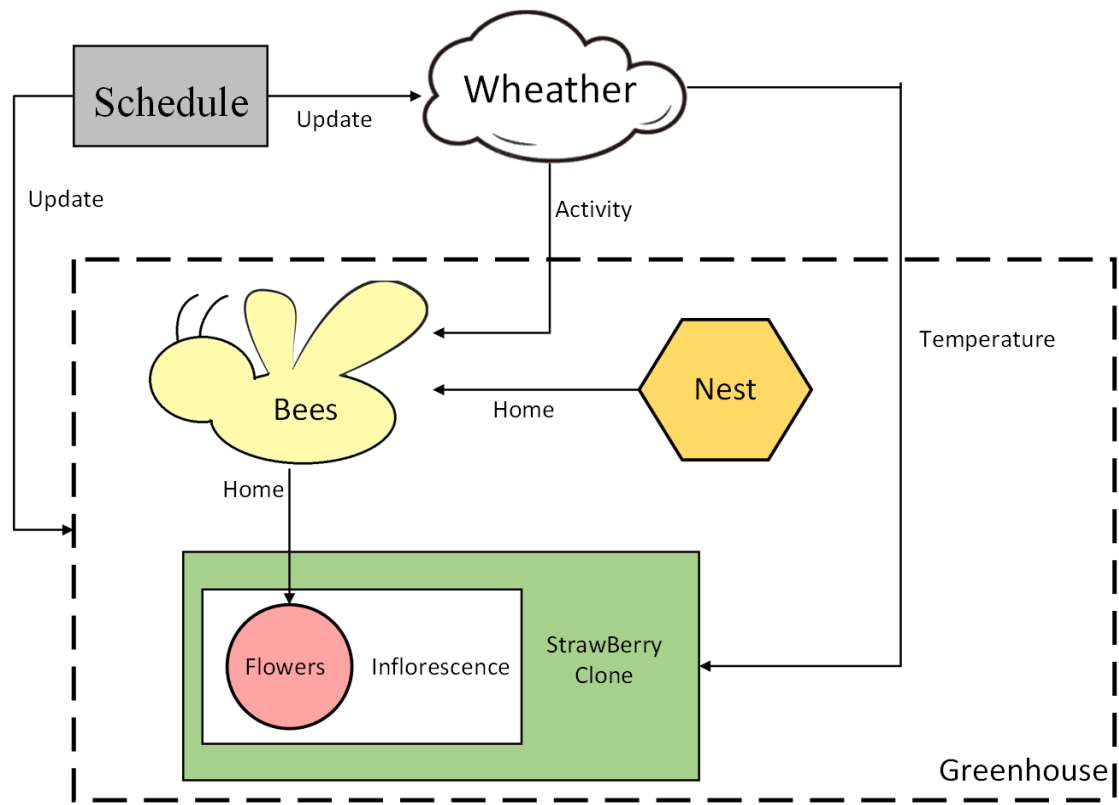

Fig.3 Entities and their interactions for strawberry pollination

The simulation covers 120 days from January to April. Then all agents are scheduled according to strawberry phenology on daily basis for strawberry development, or on minute basis for bee foraging. One simulation step is one execution in which a bee can accomplish its activity according to time and weather conditions. When the first flower blooms according to the GDD requirement, bees begin to forage from hive. We assume that a bee can accomplish one activity in 1 minute on average based on bee interaction behavior [14, 23]. We set a period of 9h for one simulated day considering bee foraging from 8:00 to 17:00. Therefore, one day contains 9*60=540 cycles in simulation. Pollination ecology data were used to parameterize the model. The simulation model incorporates much of what is known and hypothesized about the pollination ecology of a strawberry agroecosystem.

### 2.4.2 Modelling Software

The simulation model has been implemented in GAMA modelling platform [29]. The code is available from Cao or Qu upon request. The software is friendly to use and includes many adjustable parameters of strawberry phenology and bee foraging activity to adapt to variant cultivars.

The graphical user interface of the software is shown in Figure 4 and the simulated continuous strawberry growth process is shown in Figure S16.

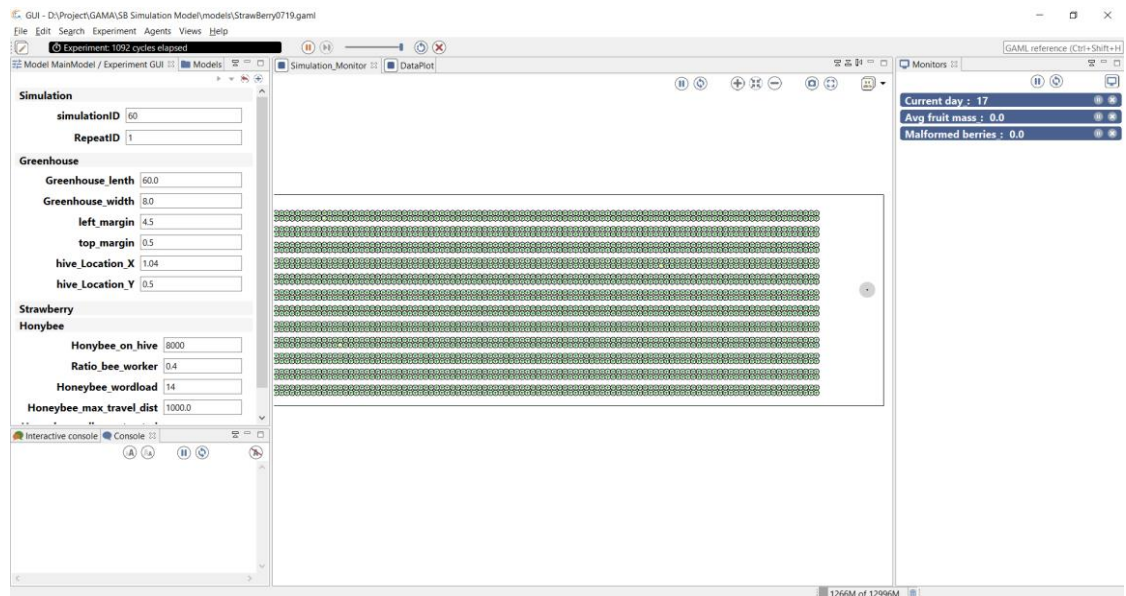

Fig.4 Friendly graphical user interface of the simulation model in GAMA

### 2.4.3 Modelling validation

The simulation results largely agreed with the existing literature [12, 15, 23, 34, 37] and the planting experience of the growers. The simulated average yield, average fruit weight and healthy fruit setting rate were basically consistent with the actual planting, which showed the validity of the model, as shown in Figure 5 and Figure 6.

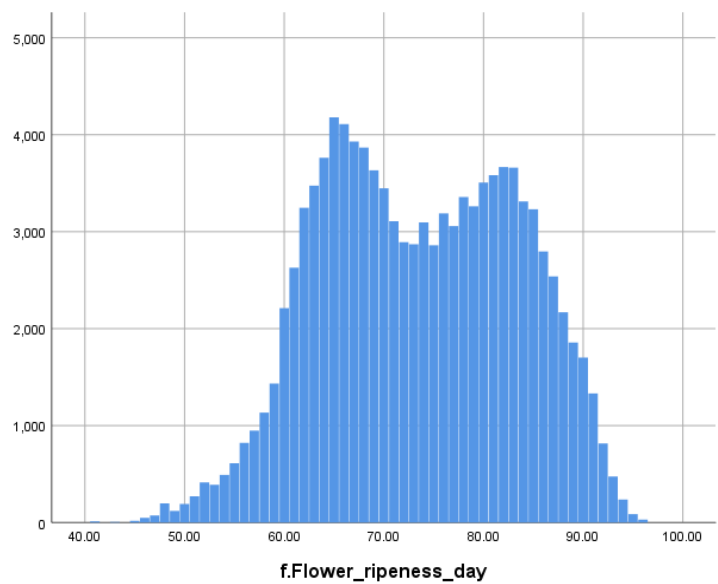

(A) Fruit setting frequency of strawberry in model simulation

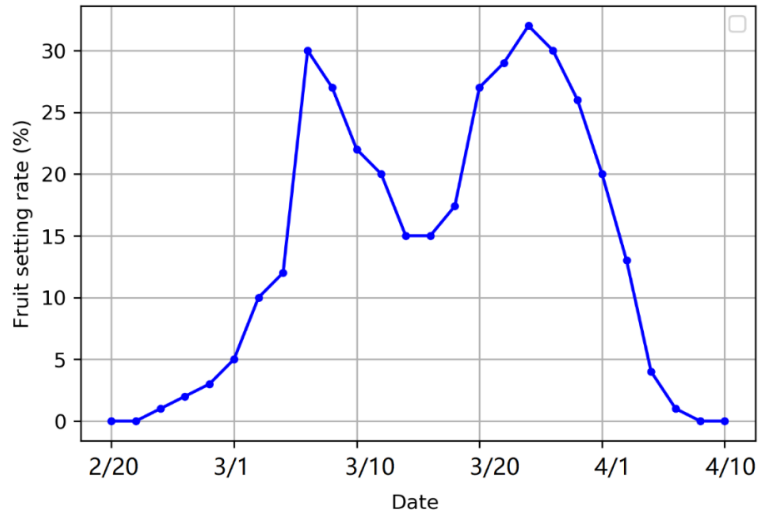

(B) Real fruit setting rate of strawberry in greenhouse planting collected from field observations

Fig.5 Simulated frequency distribution of strawberry fruiting dates in simulation and practical greenhouse planting. The X-axis represents the planting date. The Y-axis represents the corresponding flower quantity (A) and real fruit setting rate (B).

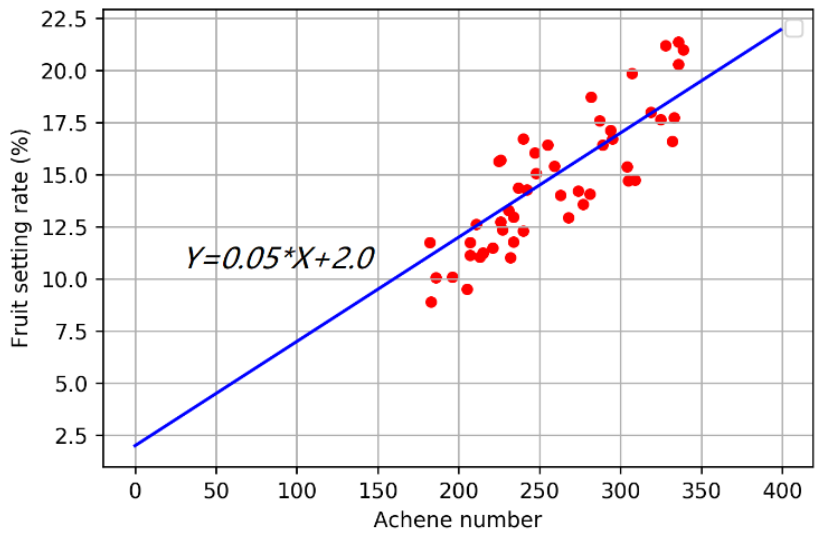

Fig.6 Blue line represents the fitted equation between achene number and fruit weight in this simulation. Red dots represent the real strawberry fruit information collected.

More detailed validation data can be obtained in the supplementary material part. These experiment results suggested our model was firmly validated, which was considered to be critical to make precise assessments and suggestions in strawberry bee pollination efficiency in a greenhouse.

# 3. Simulation experiments

Bee pollination can significantly enhance yields and increase fruit quality. However, the effects of bee density and planting pattern in a greenhouse on the fruit quality as well the causes of bee pollination advantages are largely unknown. For the first time, we analyzed in much detail the responses of fruit quality to different conditions including the bee density, pollen compatibility, stigma receptivity and diverse greenhouse planting patterns based on the firmly validated model and proposed three groups of experiments.

## 3.1 Experiment I: Bee density and fruit quality

Bee density in a greenhouse showed a strong positive correlation with fruit weight and a negative correlation with malformed fruit rate [12]. However, there is no universal conclusion about the relationship between strawberry number and bee number in the greenhouse planting. For examples, Mcgregor recommended 12–25 colonies of honeybees per hectare of strawberries for optimum fruit production [39], while Williams recommended 2.5 and Scott-Dupree 1.2 colonies of honeybees per hectare [40]. There were few studies on the effect of bee inadequacy on strawberry quality due to the high planting cost. We can use the model to test the fruit quality under different bee density conditions, and then study the influence of bee density,

Therefore, we proposed experiments to study the impact of bee density on strawberry quality in a greenhouse. A series of simulation experiments with different values of bee density was conducted with the same other factors. We tested two greenhouses of different sizes to improve generality. Based on the results, the influence of bee density can be estimated quantitatively and the optimal ratio can be selected, which may provide reasonable suggestions for growers in practical strawberry planting in a greenhouse.

## 3.2 Experiment II: Cause of bee pollination advantages

It is known that bee pollination has a significant advantage over artificial pollination. The pollination process includes accepting stage and receiving stage. Based on the planting experience and studies on bee pollination, we hypothesized that the advantage might be caused by three factors: pollen receiving rate, acceptance rate and distribution on to sigma. As for pollen distribution, some researchers believe honey bee pollination can achieve a uniform pollination of a great number of stigmata, thereby producing well-shaped and commercially high-quality fruits [30].

Alexander believed the distribution of pollen grains across the receptive stigmas seems to be important for fruit development [41]. We proposed experiments to study the impact of accepting stage and receiving stage. By adjusting the parameters of theses stages, the degree of influence can be quantitatively analyzed. Because the pollen distribution is too complex to be simulated, its impact cannot be obtained directly through this model simulation. However, the fruit quality improvement of bee pollination over artificial pollination is well known [16] and we can estimate the impact of pollen distribution indirectly by excluding the improvement caused by accepting stage and receiving stage.

Firstly, the accepting stage was studied. Since artificial pollination is mostly self-pollination in practical planting, it is greatly affected by pollen self-compatibility, while bees can pollinate between different clones, which may increase gene diversity. In the simulation, we assumed that the artificial pollination is self-pollination with genetically identical pollen. We proposed simulation experiments to test the effect of self-compatibility probability on strawberry fruit quality with a fixed bee density while keeping other factors remain unchanged. By adjusting the value of self compatibility probability in the simulation, the effect on strawberry quality was estimated quantitatively. The pollen self-compatibility probability was set to 80% preliminarily [5,37] and sensitivity analysis was conducted on it to verify its influence. Through the experimental results, we analyzed whether self-compatibility was the cause of bee pollination advantage over artificial pollination.

Secondly, the receiving stage was studied. Some research reveals that bees can lower the survival of the pollen they carry [28] and the pollen clinging to the body of honeybee foragers had a lower viability than that in the flowers. Therefore, the pollination advantages of bee pollination cannot be caused by pollen viability factor. Similarly, reference [37] shows that pollen viability has little effect on number of undeveloped fruits. Bee pollination can happen on flowers of different ages uniformly [14], while stigma viability is different at different flower ages. Normal bee pollination had been simulated in above experiments. We found artificial pollination by growers in practical strawberry plating is usually conducted within a few random days including early blooming stage (day 1-2), middle blooming stage (day 2-3) and later blooming stage (day 3-4). We proposed simulation experiments to test the effect of stigma receptivity at different flowering stages on fruit quality quantitatively. Through the experimental results, we analyzed whether stigma receptivity was the cause of bee pollination advantage over artificial pollination. According to the simulation results, we can quantify the impact of accepting stage and receiving stage on strawberry quality. It is known that bee pollination can increase the fruit weight by about 30% compared with artificial pollination.

Further, we can estimate indirectly the impact of pollen distribution factor by excluding the improvement caused by accepting stage and receiving stage. Consequently, the cause of bee pollination advantages over artificial pollination can be analyzed.

### 3.3 Experiment III: Greenhouse planting pattern and bee hive location

Normally bee hive is placed on the east side of a greenhouse in which temperature is high. This placement mode may lead to uneven pollination distribution due to the difference in the distance between the strawberry clone and bee hive. We proposed experiments to test this hypothesis.

In addition, the impact of hive location on fruit quality was studied. In the simulation, two placement modes were tested: 1) Placing one bee hive which contains all bees in different positions in a greenhouse; 2) Placing bees in two hives in different positions. Noticed that only the hive location factor was considered and the influence of sunlight on bee flight path was ignored. We proposed a hypothesis that fruit quality can be improved if bees are separated in two hives not in only one hive in greenhouse due to bee foraging behavior. To test the hypothesis, we conducted simulation experiments.

The distance between two beds is important in strawberry planting in a greenhouse. Generally, the value is set to around 0.4 m. This distance may influence the flight path of bees during foraging, thus indirectly affecting the pollination efficiency. we conducted simulation experiments to test its impact on fruit quality.

# 4. Results

### 4.1 Experiment I

Firstly, we simulated a standard greenhouse which is 80 m long and 8 m wide. In this condition, the simulation includes the growth process of 390*12*2=9,360 strawberry clones, 9,360*12=112,320 flowers and 112,320*45,000=5 billion pollen grains. It was assumed that the proportion of foraging worker bees in population is 40%. Bee density is used to describe bee abundance in a greenhouse quantitatively, i.e., the number of bees/the number of strawberry clones in the greenhouse. Through 10 simulation experiments, the effects of bee density on average strawberry healthy fruit setting rate, average berry weight, average malformed fruit rate and average yield were studied, as shown in Figure 7. More detailed simulation results are shown in the Table S1.

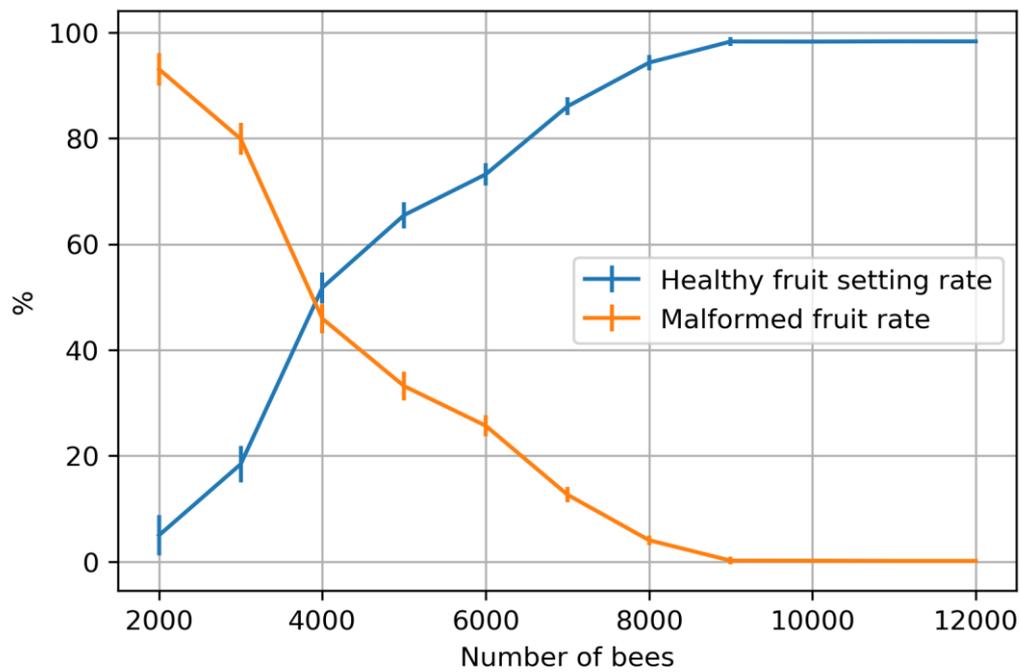

(A) Healthy fruit setting rate and malformed fruit rate data

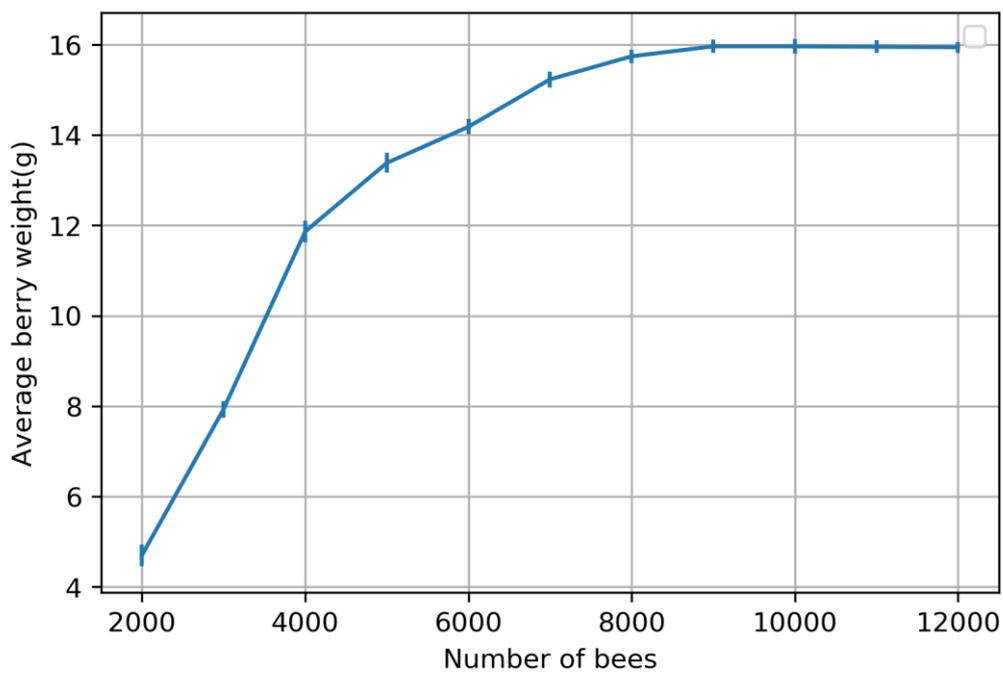

(B) Average berry weight data

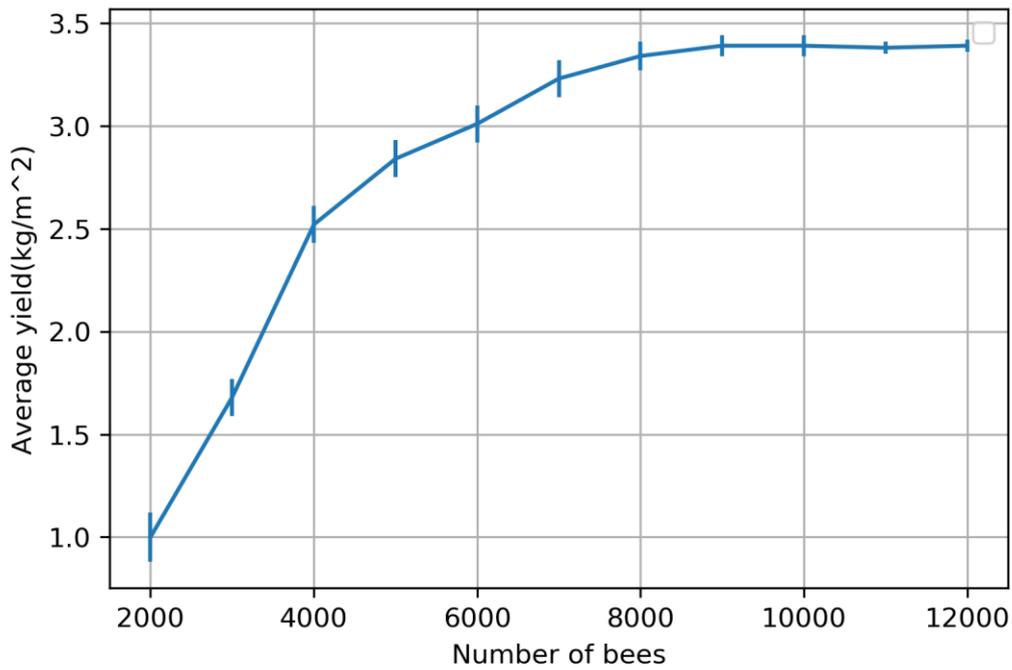

(C) Average yield data

Fig.7 Fruit quality correlated positively with the bee density in simulation

Figure 7 showed that the quality of strawberry fruit correlated positively with bee density. However, when the number of bees reached about 9,000 (bee density was about 1.00 bee/clone), this improvement was no longer significant, which showed a saturation effect on bee pollination efficiency.

Secondly, in order to further verify the simulation results, we conducted a second experiment. We simulated a small greenhouse which is 60 m long and 8 m wide. In this condition, there were 280 strawberry clones in each row, so the small greenhouse contained 280*12*2=6,720 clones. Similar simulation results were shown in Figure S8 and Table S2. When the bee density reached about 1.00 bee/clone, the improvement in fruit quality was not significant.

## 4.2 Experiment II

Firstly, the effect of the pollen compatibility factor on strawberry quality was studied. We conducted two experiments fixing the number of bees to 8000 and 4000 respectively while keeping other factors remaining unchanged. The simulation with 8000 bees represents the scenario in which bees are adequate. The results for this condition are shown in Table 5.

Table 5. Impacts of pollen compatibility in simulation with 8000 bees

| Pollen compatibility | Average berry weight (g) | Malformed fruit rate (%) |
|---|---|---|
| 0.40 | 15.72 ± 0.12a | 3.88 ± 0.43a |
| 0.50 | 15.76 ± 0.19a | 3.13 ± 0.81a |
| 0.60 | 15.78 ± 0.23a | 3.50 ± 0.44a |
| 0.70 | 15.76 ± 0.17a | 4.18 ± 0.47a |
| 0.80 | 15.74 ± 0.24a | 4.07 ± 0.86a |
| 0.90 | 15.81 ± 0.12a | 3.31 ± 0.74a |
| 1.00 | 15.77 ± 0.11a | 3.83 ± 0.91a |

Means ± standard error (n=10) with each column followed by the same letter are not significantly different (P>0.05)

The data presented in Table 5 shows that there was no significant difference between groups (P>0.05). It suggests that self-compatibility factor may have little impact on the strawberry fruit quality. In order to further verify the results, we reduced the number of bees to 4000 to test the scenario in which bee abundance is insufficient. The results are shown in Table 6 and are basically similar to those in Table 5.

Table 6. Impacts of pollen compatibility in simulation with 4000 bees

| Pollen compatibility | Average berry weight (g) | Malformed fruit rate (%) |
|---|---|---|
| 0.40 | 11.69 ± 0.32a | 46.54 % ± 1.76a |
| 0.50 | 11.95 ± 0.29a | 47.77 % ± 1.30a |
| 0.60 | 11.98 ± 0.26a | 46.71 % ± 1.41a |
| 0.70 | 11.96 ± 0.34a | 48.63 % ± 2.61a |
| 0.80 | 11.86 ± 0.31a | 45.93 % ± 2.56a |
| 0.90 | 11.86 ± 0.19a | 47.95 % ± 2.47a |
| 1.00 | 11.81 ± 0.39a | 45.65 % ± 2.40a |

Means ± standard error (n=10) with each column followed by the same letter are not significantly different (P>0.05)

Secondly, the effect of stigma receptivity factor on strawberry quality was studied. We conducted experiments to simulate the bee pollination and artificial pollination process in different flower ages. As for artificial pollination, the average proportion of received pollen with high viability is about 0.99, 0.93, and 0.73 in early blooming stage, middle blooming stage and later blooming stage, respectively according to the above formula. It should be noticed that the artificial pollination process is influenced by compatibility factor. The results are shown in Table 7. The data presented shows that there were significant differences in the fruit quality of simulation with different stigma receptivity values (P<0.05). It suggested that the earlier the pollination process is in the blooming period, the better the fruit quality will be. However, this improvement is not huge quantitively. The average fruit weight in early blooming stage is less than 5% higher than the that in later blooming stage.

Table 7. Impacts of stigma receptivity

| Stigma receptivity | Average berry weight (g) | Malformed fruit rate (%) | Pollination pattern |
|---|---|---|---|
| $R = e^{-0.01*age^{3.6}}$ | 15.74 ± 0.24a | 4.07 ± 0.86a | Bee pollination |
| 0.99 | 15.93 ± 0.11b | 2.27 ± 0.67b | Early stage |
| 0.93 | 15.89 ± 0.13b | 3.52 ± 0.98c | Middle stage |
| 0.73 | 15.71 ± 0.14a | 4.24 ± 0.82a | Later stage |

Means ± standard error (n=10) with each column followed by the same letter are not significantly different (P>0.05)

## 4.3 Experiment III

Firstly, the relationship between strawberry clone coordinate in a greenhouse and fruit quality was studied. The results based on SPSS are shown in Figure 8. More detailed results are shown in Figure S14 and Figure S15. The data presented in Figure 8 show that the fruit weight distribution on X-coordinate is not uniform with a significant difference (P < 0.05). As expected, fruit weight distribution on Y-coordinate was basically uniform without significant difference (P > 0.05).

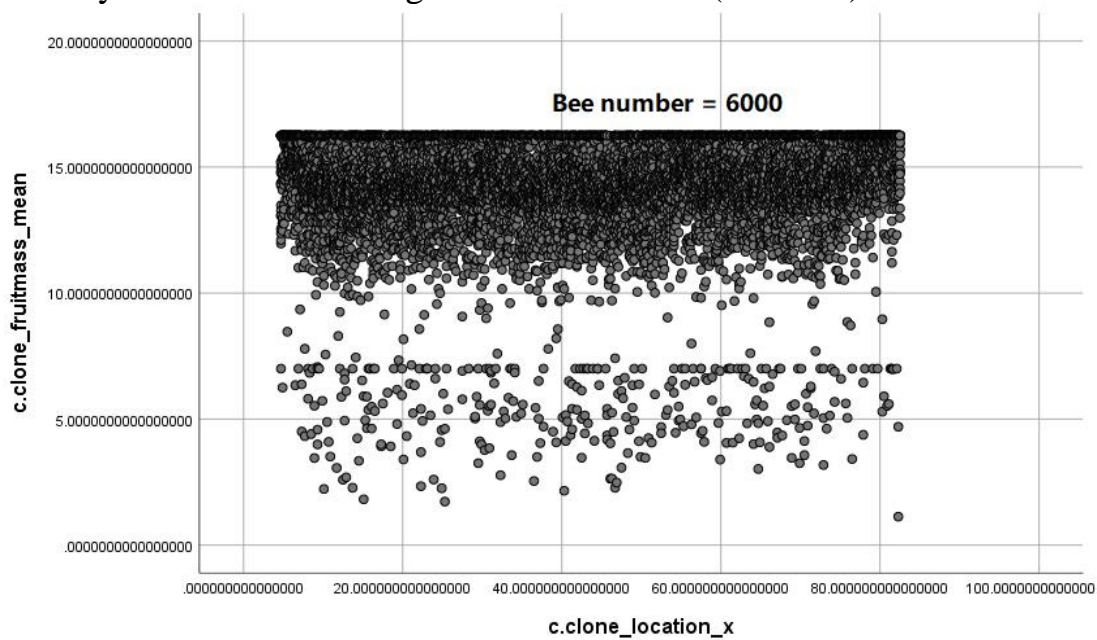

Fig.8 Diagram based on SPSS to show the relationship between strawberry clone X-coordinate and fruit quality. High X-value represents short distance to the hive located in east side of greenhouse.

Secondly, the impact of hive location on fruit quality was studied. We placed all bees in one hive and then placed the hive in different location in a greenhouse. The four sides and center were tested. The results are shown in Table 8. The data presented shows that there was no significant difference (P<0.05) in fruit quality and the hive location may have little effects on fruit quality with only one hive in a greenhouse.

Table 8. Impacts of diverse hive locations with only one hive

| Location of hive | Average berry weight(g) | Malformed fruit rate (%) |
|---|---|---|
| Middle of east side | 15.74 ± 0.24a | 4.07 ± 0.86a |
| Middle of west side | 15.82 ± 0.12a | 3.56 ± 0.91a |
| Greenhouse center | 15.72 ± 0.16a | 3.74 ± 0.79a |
| Middle of south side | 15.75 ± 0.13a | 4.18 ± 1.34a |
| Middle of north side | 15.78 ± 0.14a | 3.71 ± 0.74a |

Means ± standard error (n=10) with each column followed by the same letter are not significantly different (P>0.05)

Then we placed bees in two hives to test pollination efficiency in this condition. As discussed above, when the bee number reached around 8000, the improvement of fruit quality from bee foraging behavior was not significant due to a saturation effect. As a result, the bee number was set to 4000 in this simulation. There were 2000 bees in one hive on the east side and 2000 bees in one hive on the west side in a greenhouse. The results are shown in Table 9. The data presented shows that there was a significant difference in fruit quality (P<0.05).

Table 9. Location impacts of two hives

| Location of hive | Average berry weight(g) | Malformed fruit rate (%) |
|---|---|---|
| Only east side | 11.86 ± 0.31a | 45.93 % ± 2.56a |
| West side + east side | 12.42 ± 0.33b | 43.37% ± 2.70b |

Means ± standard error (n=10) with each column followed by the same letter are not significantly different (P>0.05)

Thirdly, the effect of adjusting the bed spacing on the fruit quality was studied by simulation. Due to the irrigation demand and the limitation of greenhouse area, the distance was set to 0.35m, 0.40m and 0.45m for comparison. The results are shown in Table 10. The simulation results showed there was no significant difference in berry weights (P>0.05) and a significant difference in malformed fruit rates (P<0.05). It suggested that bed spacing may have little effects on strawberry fruit weight and moderate effects on malformed fruit rate.

Table 10. Impacts of the distance between beds

| Distance between beds | Average berry weight (g) | Malformed fruit rate (%) |
|---|---|---|
| 0.35 m | 15.77 ± 0.15a | 3.32 ± 0.86a |
| 0.40 m | 15.74 ± 0.12a | 4.07 ± 0.89b |
| 0.45 m | 15.72 ± 0.16a | 4.17 ± 0.85b |

Means ± standard error (n=10) with each column followed by the same letter are not significantly different (P>0.05)

## 5. Discussion and Conclusions

In experiment 1, the results showed that inadequate bees in the greenhouse have a significant adverse effect on the fruit quality. As a result, sufficient bee density is a must for strawberry planting. The data presented shows that when the bee density was higher than 1.00 bee/clone, the fruit quality was not significantly improved. There were similar experimental phenomena in two greenhouses with different sizes in the simulation. We suggested that it is a saturation effect on bee pollination efficiency. In practical greenhouse planting, strawberry pollination may be affected by mechanical actions, diseases and other factors, with a result that it is necessary for growers to ensure the bees outnumber the strawberry clones (bee density is more than 1.00 bee/clone) in a greenhouse.

In experiment 2, by comparing the effects of pollen compatibility on fruit quality in two scenarios in which bees are sufficient and insufficient, we suggested that pollen compatibility is not the main cause of pollination advantage of bee pollination, and pollen self-compatibility factor may have little effect on fruit quality (P>0.05) in practical strawberry planting. We analyzed that it is because: 1) The number of pollen grains removed by a bee per visit is high, while the number of pollen grains deposited by a bee per visit is low, providing a small opportunity for a flower to accept the pollen produced by the same clone; 2) Strawberry clones are close to each other in a greenhouse, resulting in a high probability of bees pollinating between different strawberry clones and flowers. Therefore, the probability of a flower accepting strawberry pollen from the same clone is low, leading to the little impact of pollen compatibility on fruit quality in strawberry planting. In addition, we found that only about 4.2% of the pollen came from the same strawberry clone, and most of the pollen came from other clones through the analysis of the simulation results, which validated our hypothesis. Therefore, we suggested that self-compatibility is not the cause of bee pollination advantage over artificial pollination.

The data in Table 7 showed that stigma receptivity is one factor causing the bee pollination advantage. We suggested that strawberry growers should conduct artificial pollination in time after blooming if they do not use bee pollination. However, quantitative analysis showed the berry weight improvement rate is less than 5% which is far lower than practical planting [16,17]. It revealed that stigma receptivity (artificial pollination time schedule) is a cause but not the primary cause of bee pollination advantage over artificial pollination. Based on the above discussion, we believed that bees help distribute pollen grains to all strawberry pistils, promoting well-shaped fruits, which is the primary cause of bee pollination advantage over artificial pollination in strawberry planting. It can be explained by an interactive effect of even pollen distribution and meeting a threshold of pollen grains per flower. In addition, the even pollen distribution can effectively reduce the malformed fruit rate. Therefore, in bee pollination, not only the pollen transport but the even pollen distribution caused by bee behavior are significantly important.

To conclude, the results in experiment 2 suggested that the even distribution of pollen in pistil during bee pollination is the primary cause and stigma receptivity is a secondary cause of advantage for bee pollination over artificial pollination.

In experiment 3, it revealed that fruit weight distribution was non-uniform on X-coordinate but basically uniform on Y-coordinate according to the distance from the hive. We thought that the phenomenon was due to the foraging distance constraints of bees [14] because the distance influenced the bee flight path when foraging. In addition, we found that placing bees in two hives of different locations in a greenhouse can improve strawberry fruit quality, which suggested a novel effective method for growers to place the bee hive. Strawberry growers can place bees in multiple hives not only one hive and then place these hives in different locations in a greenhouse. In this way, the influence of bee foraging distance constraints can be reduced. These results in experiment 3 also suggested that when the bed spacing was appropriately reduced, the malformed fruit rate slightly declined, providing a suggestion that the growers can appropriately reduce the bed spacing according to the actual greenhouse environment without interference in planting.

In conclusion, we modelled the greenhouse spatial and temporal structure of strawberry plants and honey bee foraging behavior at the individual bee, flower, inflorescence, and clone level. The interactions between strawberry clones and bees are scheduled by phenology and physiological conditions. This model is based on a variety of previous strawberry research and provides a software model with many adjustable parameters that allow users to adapt to different cultivars and planting environments. This model allows users to explore how various factors, including bee density, pollen compatibility, pollen viability, stigma receptivity, bed spacing, hive location, changes in weather conditions and foraging behavior as well as greenhouse environment influence the pollination process and fruit quality, particularly healthy fruit setting rate, average berry weight, malformed fruit rate and average yield. The graphical user interface of software lowers barriers to using simulation for researchers or interested growers with little or no experience in modelling, and enables them to test hypotheses, develop theories and assess strawberry management strategies. Based on this firmly validated model, we proposed some planting suggestions for strawberry growers.

Despite the validation, this model still has some limitations. As shape of the fruit was observed with the naked eye by researcher and fruit malformation can be caused by diverse factors, there may be discrepancies in the malformed fruit rate between simulation and empirical data [16, 36, 37]. In addition, the influence of sunlight, humidity and other environment factors on bee foraging behavior was ignored in the simulation. More attention shall be paid to these limitations in further study.

# Declaration of Competing Interest

The authors declare that they have no known competing financial interests or personal relationships that could have appeared to influence the work reported in this paper.

# Acknowledgements

The authors acknowledge funding received from the Key Research Program of the Science Foundation of Shandong Province (ZR2020KE001) and this is also a publication of the Enroll Plan of Young Innovative Talents of Shandong Province (Big Data and Ecological Security Research and Innovation Team Project).